# Ambipolar bistable switching effect of graphene

Young Jun Shin,[1,2] Jae Hyun Kwon,[1,2] Gopinadhan Kalon,[1,2] Kai-Tak Lam,[1] Charanjit S. Bhatia,[1] Gengchiau Liang,[1] and Hyunsoo Yang [1,2,a)]

[1]*Department of Electrical and Computer Engineering, National University of Singapore, 117576, Singapore*

[2]*NUSNNI-NanoCore, National University of Singapore, 117576, Singapore*

Reproducible current hysteresis is observed in graphene with a back gate structure in a two-terminal configuration. By applying a back gate bias to tune the Fermi level, an opposite sequence of switching with the different charge carriers, holes and electrons, is found. The charging and discharging effect is proposed to explain this ambipolar bistable hysteretic switching. To confirm this hypothesis, one-level transport model simulations including charging effect are performed and the results are consistent with our experimental data. Methods of improving the ON/OFF ratio of graphene resistive switching are suggested.

[a)] e-mail address: eleyang@nus.edu.sg

Graphene, a flat monolayer of carbon atoms packed into a honeycomb lattice, has attracted considerable attention due to its peculiar and superior characteristics.[1-3] Various prototypes of applications have been demonstrated utilizing its unique physical properties.[4, 5] Recently, 100 GHz graphene transistors were successfully fabricated, indicating that graphene can be a promising candidate for the post-silicon era.[6] Prototypes of memory devices have been fabricated with graphitic materials such as graphite, carbon nanotube, and graphene oxide by taking advantage of the current hysterisis instead of the conventional memory structures, such as those used in the flash or dynamic random access memories.[7-9] However, current hysteresis or resistive switching has not been explored in pristine graphene and metallic carbon nanotube due to the absence of a band gap.

In this study, current hysteresis in graphene devices is investigated in a two-terminal configuration showing bistable states with a back gate bias to tune the Fermi level of graphene. Reproducible current hysteresis is still observed after passivating graphene with a 2 nm thick $Al_2O_3$ layer. We observed that the switching sequence of the current hysteresis of p-type graphene is inverted to that of n-type graphene. To explain the observed ambipolar bistable switching effect, we propose the charging and discharging effect (CDE). Quantum transport simulations, self-consistently coupled with charging energy calculation[10], are performed and the results indicate that CDE plays an important role in hysteretic switching. In addition, resistive switching of graphene devices is found to be stable over more than 100 cycles of operation. To improve the ON/OFF ratio of the resistive switching device, various methods, such as band gap engineering and chemical doping, are suggested. By employing this unique ambipolar switching effect, multi-state devices based on graphene are proposed.

Single and multi-layer graphene are prepared by micromechanical exfoliation and transferred to a highly p-doped Si substrate, which is covered by a layer of 300 nm thick $SiO_2$. Mechanically cleaved graphene is identified by optical microscopy and Raman spectroscopy. Further details on the sample preparation and identification can be found in our earlier work.[11] The electrodes are patterned using standard lithography and Cr/Au (5 nm/80 nm) is deposited by a thermal evaporator, followed by standard lift-off procedures. For passivation, a 2 nm thick Al layer is deposited and the sample is then exposed to air for natural oxidation. The complete oxidation of Al is verified via X-ray photoelectron spectroscopy. The *I-V* characteristics of graphene with a two-probe configuration are measured under ambient conditions. To apply a back gate bias, the cathode is connected to the back gate and the leakage current through the $SiO_2$ layer is monitored.

Figure 1(a) shows the Raman spectra of single and multi-layer graphene samples. From the appearance of in-plane vibrational *G* band (1581 $cm^{-1}$) and two-phonon *2D* band (2680 $cm^{-1}$), single layer graphene can be clearly identified. The number of layers of the graphene sample is determined by measuring the width of the *2D* band.[12] The inset of Fig. 1(a) shows the optical image of graphene on top of a Si wafer covered with 300 nm thick $SiO_2$ after the deposition of Cr/Au contacts. Figure 1(b) shows the resistance as a function of the back gate voltage. Graphene is found to be p-type without an external electric field, which is seen as a shift of the Dirac point to a positive back gate voltage. The Dirac peak shift can be attributed to unintentional doping such as water molecules.[13] In typical transport experiments of graphene, applying a high voltage between the source and drain is not preferred, as this causes a device breakdown. Some experiments, which applied a high voltage (> 2 V) to graphene devices, reported nonlinear *I-V* characteristics similar to our *I-V* curves in the insets of Fig. 1(b).[14-16] The difference in the maximum current density between the forward and backward sweeps is ~ $0.5 \times 10^7$ A/cm$^2$ with a sample width of 4 µm and a thickness of 0.35 nm at ± 4 V. In addition to the nonlinear *I-V* curves, current hysteresis is observed as shown in Fig. 1(c-f) at different back gate bias. The *I-V* characteristics exhibit a typical unipolar (or symmetric) switching behavior in which the switching procedure does not depend on the polarity of the voltage and current signal. Moreover, we find an interesting phenomenon in which the sequence of switching is reversed when the charge carrier type is changed by an external bias. As shown in Fig. 1(c-d), switching starts from a low resistance state and ends up with a high resistance state for both the positive and negative sweeps in the hole transport regime. On the other hand, when the charge carrier becomes electrons as shown in Fig. 1(e-f), switching starts from a high resistance state and ends up with a low resistance state for both the polarity sweeps. Ambipolar characteristics at different back gate biases imply that intrinsic physical properties of graphene are responsible for the observed current hysteresis.

      Figure 2(a) shows the resistance versus back gate voltage of the passivated sample. Without an external bias, the charge carriers of graphene are holes. Two types of current hysteresis depending on the type of carriers are still observed despite the devices are completely covered by a 2 nm thick $Al_2O_3$ layer. However, the current hysteresis becomes more complicated with the additional oxide layer. Switching behavior, in both hole and electron regimes for the negative bias voltage sweep, is found to be similar to

the switching characteristics of graphene devices without the oxide layer. However, in the case of hole transport in Fig. 2(b), for the positive bias voltage sweep, the current hysteresis starts from a high resistance state and changes to a low resistance state before the voltage between the source and drain is swept backward. When the voltage is swept backward from 5 V, the system becomes a high resistance state again, and before the polarity of sweep voltage becomes negative, it turns to a low resistance state. This observation demonstrates the importance of the graphene surface for current hysteresis. Nonetheless, the interesting point is that the sequence of current hysteresis from p-type graphene in Fig. 2(b) is exactly opposite to that of n-type graphene in Fig. 2(c). More than 15 devices, without and with an oxide layer, are tested and all of these samples show current hysteresis as presented in Fig. 1 and 2 depending on whether it is fabricated without or with an oxide layer, respectively. In addition, amorphous glassy carbon films of 2 nm thick by pulse laser deposition are measured to examine any hysteresis switching behavior, shown in Fig. 2(d). Thin Au strips without graphene in the lower inset of Fig. 2(d) are also tested to check the possible switching effect due to electrical annealing and residual $TiO_x$.[17] No hysteretic switching is found and linear *I-V* characteristic is observed for both cases, demonstrating that our observation from graphene devices is unique.

In order to verify that CDE is the cause of hysteretic resistive switching, we perform *I-V* measurements in both vacuum ($10^{-8}$ Torr) and ambient conditions. The hysteresis is much weaker under vacuum conditions due to the lack of the charging sources as shown in Fig. 3(a). The possible sources of CDE are: 1) unintentional particles between contacts and graphene, which can be introduced during or before the deposition of contacts; 2) particles between graphene and the substrate; and 3) dangling bonds from the edge of graphene. Two opposite types of CDE sources can be assumed. First, when a bias voltage is applied, the positively charged hydrogens ($H^+$) or similar polarity groups are removed in the case of p-type graphene, but attached to the n-type graphene surface. Conversely, when a bias voltage is applied, the negatively charged hydroxyl ions ($OH^-$) or similar polarity groups are detached from n-type graphene, but attached to p-type graphene. The charging energy in p-type graphene increases with bias voltage, which requires a larger change in bias for the same electron conduction, therefore, the current decreases when the bias voltage changes from positive to zero. On the other hand, for the n-type devices, the increase of bias voltage leads to a decrease of the charging energy and an increase of the current in the backward sweep. The difference in current is getting bigger as the range of sweep voltage increases shown in Fig. 3(b). This implies that more charging sources become activated with a higher bias, which leads to a larger hysteresis curve.

In order to understand CDE, a two-terminal one-level transport model where the carriers flow across the channel via an isolated energy state, is employed, with the device potential obtained self-consistently with charging energy calculation[10, 18]. In this model, the lowest unoccupied molecular orbital (LUMO) conduction or the highest occupied molecular orbital (HOMO) conduction can be considered as n-type or p-type conduction channels, and they are set at 0.2 and -0.2 eV, respectively. Furthermore, the carrier escape rate between the contact and the channel is set to $5\times10^{-3}$ $s^{-1}$ at 300 K. A schematic of the model for HOMO channel conduction is shown in the lower inset of Fig. 3(c). As the drain bias increases, the chemical potential of the drain moves downwards, and when it is lower than the energy of the channel state, a net current occurs. The charge occupation of the channel state changes from filled to partially-filled and the charging energy is thus changed. The simulated *I-V* characteristics, as shown in Fig. 3(c), present the hysteresis behavior due to a change in the charging energy as the bias voltage increases. Moreover, the different patterns of n-type and p-type devices exactly match with the experimental data shown in the insets of Fig. 1(b). Figure 3(c) also shows that the simulated hysteresis loop of the p-type device becomes larger, as the sweeping bias increases thus the difference in charging energy increases. Although our model does not account for the complex switching behavior shown in Fig. 2(b-c), the model clearly reveals the physical origin of the hysteresis switching for p- and n-type graphene devices.

The reproducibility of switching effect is tested. Input voltage is repeated in the following sequence: 0V, 4V, 8V, 4V, and 0V, and the output current is measured at 4 V for the forward and backward sweeps. The resistance change ratio, $\Delta R/R = (R_{OFF}-R_{ON})/R_{ON}$ is around 13% in Fig. 3(d). Even though the ratio of resistance change is small, 100 cycles of operation can be repeated without serious degradation. Yao *et al.* engineered two-terminal nonvolatile memory with single-walled carbon nanotube, and reported that current hysteresis was not observed for metallic carbon nanotube due to the absence of a band gap.[8] As the electrical characteristic of graphene is similar to that of a metallic system, it is difficult to make a large ON/OFF ratio from graphene, which is a major hurdle for graphene devices.[19] However, by patterning graphene into nano-ribbons, a band gap can be induced, resulting in a higher ON/OFF ratio.[20, 21] Furthermore, the band gap of bilayer graphene can be opened up with a gate bias[22], and metallic graphene can be turned into an insulator by chemical doping.[4] As a result, graphene device is expected to take advantage of its high mobility for fast resistive switching memory applications.[6] In addition, by utilizing the observed ambipolar switching effect, graphene devices can be set at multiple states depending on the gate bias voltage, leading to a possibility of multi-bit memory devices.

In summary, the study of reproducible current hysteresis in graphene is presented. We observe that the sequence of hysteresis switching with different type of the carriers, n-type and p-type, is inverted and we proposed that CDE is responsible for the observed ambipolar switching effect, supported by quantum simulations. In addition, band gap engineering is proposed to improve the ON/OFF ratio of resistive switching and multi-bit memory devices can be realised based on the observed hysteresis switching. Our observation demonstrates an opportunity to realize two-terminal memory devices with well studied three-terminal FET graphene devices on a same graphene based platform if the ON/OFF ratio is improved.

Figure Captions

Figure 1. (a) Raman spectra of single layer and multi-layer graphene. The inset in (a) shows an optical image of a device (scale bar is 3 µm). (b) Resistance vs. back gate voltage ($V_g$) of a device. The upper and lower insets in (b) show typical *I-V* data in p-type ($V_g = 70$ V) and n-type ($V_g = 130$ V) devices. (c-f) Resistance vs. bias voltage at different $V_g$.

Figure 2. (a) Resistance vs. $V_g$ of a device with 2 nm thick $Al_2O_3$. Typical *I-V* characteristics in p-type graphene (b) and n-type graphene (c). (d) *I-V* data of glassy carbon film. The upper inset shows the Raman spectra of glassy carbon and the lower inset shows *I-V* curve of a Au strip.

Figure 3. (a) *I-V* data of p-type graphene in both vacuum and air without a gate bias. (b) *I-V* with different voltage sweep ranges. (c) The simulated *I-V* of a p-type graphene device. The upper inset shows simulated *I-V* of a n-type graphene device. The lower inset represents the one-level model used for this simulation with $\mu_S$ and $\mu_D$ being the chemical potentials of the source and drain, $\varepsilon$ is the energy of the conduction state in the channel and the shaded regions are filled with electrons. (d) 100 cycles of ON/OFF switching.

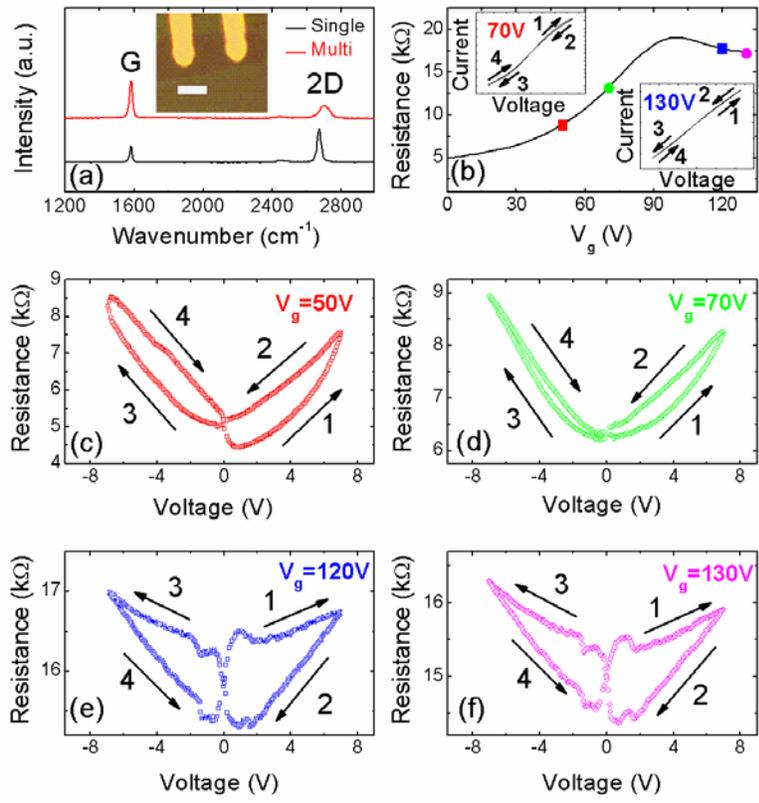

Figure 1.

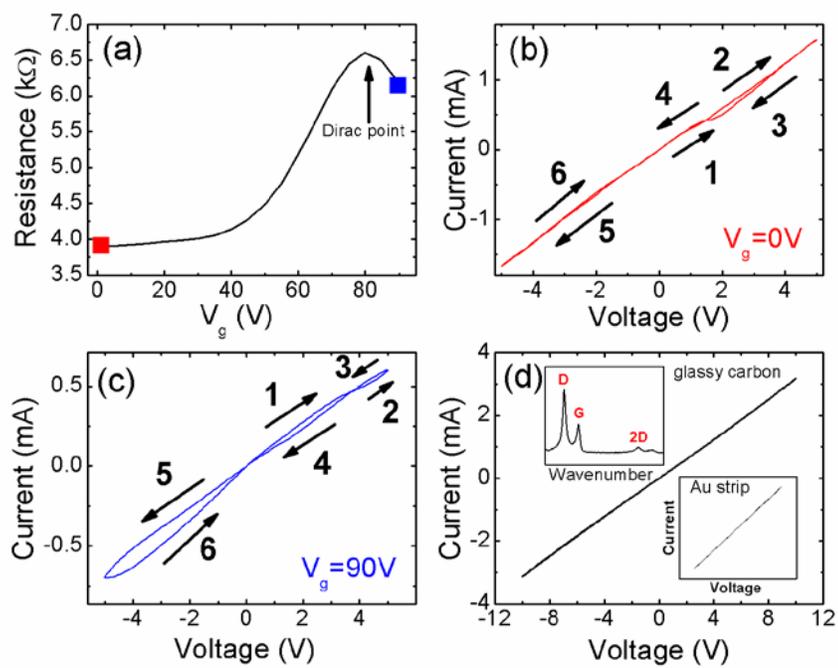

Figure 2.

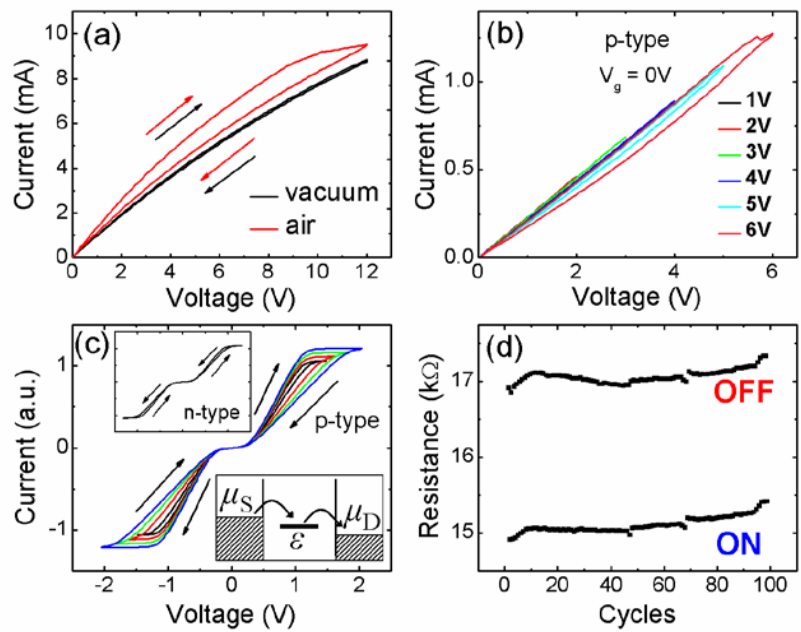

Figure 3.